# Classifying induced superconductivity in atomically thin Dirac-cone materials

**Evgueni F. Talantsev**[1,2,*]

[1] M. N. Miheev Institute of Metal Physics, Ural Branch, Russian Academy of Sciences, 18, S. Kovalevskoy St., Ekaterinburg, 620108, Russia; evgeny.talantsev@imp.uran.ru
[2] NANOTECH Centre, Ural Federal University, 19 Mira St., Ekaterinburg, 620002, Russia; evgeny.talantsev@urfu.ru
[*] Correspondence: evgeny.talantsev@imp.uran.ru; Tel.: +007-912-676-0374

**Abstract:** Recently, Kayyalha *et al.* (*Phys. Rev. Lett.*, **2019**, *122*, 047003) reported on anomalous enhancement of the self-field critical currents, $I_c(sf,T)$, at low temperatures in Nb/BiSbTeSe$_2$-nanoribbon/Nb Josephson junctions. The enhancement was attributed to the low-energy Andreev bound states arising from winding of the electronic wave function around the circumference of the topological insulator BiSbTeSe$_2$ nanoribbon. In this paper, we show that identical enhancement in $I_c(sf,T)$ and in the upper critical field, $B_{c2}(T)$, at approximately same reduced temperatures, were reported by several research groups in atomically thin junctions based on a variety of Dirac-cone materials (DCM) earlier. Our analysis shows that in all these S/DCM/S systems the enhancement is due to a new superconducting band opening. Taking in account that several intrinsic superconductors also exhibit the effect of new superconducting band(s) opening when sample thickness becomes thinner than the ground state out-of-plane coherence length, $\xi_c(0)$, we strength our previous proposal that there is a new phenomenon of additional superconducting band(s) opening in atomically thin films.

## 1. Introduction

Intrinsic superconductors can be grouped in 32 classes under "conventional", "possibly unconventional" and "unconventional" categories according to the mechanism believed to give rise to superconductivity [1]. Despite some differences, all intrinsic superconductors can induce superconducting state in non-superconducting materials by the Holm-Meissner effect [2] (also designates as the proximity effect [3,4]). As direct consequence of this, non-dissipative transport current can flow throw the non-superconducting material in superconductor/non-superconductor/superconductor (S/N/S) junctions. The amplitude of this non-dissipative transport current at self-field conditions (when no external magnetic field is applied), $I_c(sf,T)$, was given by Ambegaokar and Baratoff (AB) [5,6]:

$$I_c(sf,T) = \frac{\pi \cdot \Delta(T)}{2 \cdot e \cdot R_n} \cdot tanh\left(\frac{\Delta(T)}{2 \cdot k_B \cdot T}\right), \tag{1}$$

where $\Delta(T)$ is the temperature-dependent superconducting gap, $e$ is the electron charge, $R_n$ is the normal-state tunneling resistance in the junction, and $k_B$ is the Boltzmann constant.

Many interesting physical effects are expected if non-superconducting part of S/N/S junction will be made of single-layer graphene (SLG) [7], multiple-layer graphene (MLG) [8], graphene-like materials [9], and many other new 2D- and nano-DCMs which are under on-going discover/invent/exploration stage now [10-39]. One certainly interesting class of S/N/S junctions is when non-superconducting part of the device made of topological insulators (TI) [40-47]. Temperature dependent self-field critical currents, $I_c(sf,T)$ in this class of junctions were first reported by Veldhorst *et al.* in Nb/Bi$_2$Te$_3$/Nb [18], and later by Kurter *et al.* in Nb/Bi$_2$Se$_3$/Nb [19], by Charpentier *et al.* in Al/Bi$_2$Te$_3$/Al [42], and by other research groups in different systems (extended reference list for studied S/TI/S junctions can be found in Refs. 45,46).

Recently, Kayyalha *et al.* [48] report on anomalous enhancement of $I_c(sf,T)$ in Nb/BiSbTeSe$_2$-nanoribon/Nb junction at temperatures of $T \leq 0.25 \cdot T_c$. They confirmed the effect in all five studied junctions [48], for which TI parts were made of BiSbTeSe$_2$ flakes with thicknesses, $2b$, varied from 30 nm to 50 nm, and flakes widths, $2a$, varied from 266 nm to 390 nm. We note, that in all these S/TI/S junctions, BiSbTeSe$_2$-nanoribbons thicknesses and widths were smaller than the ground state superconducting coherence length, $2b \ll 2a < \xi(0) \sim 600$ nm in these devices [48]. For one junction, made of wider BiSbTeSe$_2$-nanoribbon, $2a = 4$ μm (Fig. S4 of Supplementary Information of Ref. [48]), measurements were performed only at low temperatures, $T < 2$ K, which is about $T < 0.2 \cdot T_c$ (if we take in account, that Nb has $T_c = 8.9$-$9.6$ K [49]), and, thus, more experimental studies are required for this 4-μm wide Nb/BiSbTeSe$_2$-nanoribbon/Nb junction to see the $I_c(sf,T)$ enhancement.

Here we need to stress, that identical $I_c(sf,T)$ enhancement (or, in another words, $I_c(sf,T)$ upturn [12]) at approximately the same reduced temperature of $T \leq 0.25 \cdot T_c$ in atomically-thin S/N/S junction was first reported by Calado *et al.* [12] in MoRe/SLG/MoRe junction in 2015. One years later, less prominent $I_c(sf,T)$ enhancement (however, which is still very clearly visible in raw experimental data [50]), in nominally the same MoRe/SLG/MoRe junctions at $T \leq 0.25 \cdot T_c$ was reported by Borzenets *et al.* [15]. Based on this, it will be incorrect to attribute the $I_c(sf,T)$ enhancement at low reduced temperatures in Nb/BiSbTeSe$_2$-nanoribbon/Nb [48] to unique property of S/TI/S junctions.

In addition, this is important to mention that Kurter *et al.* [19] were the first who reported $I_c(sf,T)$ enhancement in S/TI/S junction at reduced temperature of $T \leq 0.25 \cdot T_c$. In their Nb/Bi$_2$Se$_3$/Nb junctions, Bi$_2$Se$_3$ flake has thickness of $2b = 9$ nm, and, thus, the condition of $2b < \xi_c(0)$ was also satisfied.

In overall, as S/TI/S [19,48], as S/SLG/S [12,15], studied junctions, for which the effect of the low-temperature $I_c(sf,T)$ enhancement was observed have non-superconducting parts thinner than the ground state out-of-plane coherence lengths, $\xi_c(0)$. Truly, SLG thickness is $2b = 0.4$-$1.7$ nm [50] and thus the condition of $2b \ll \xi_c(0)$ satisfies for any SLG-based junctions.

We have to note that several intrinsic superconductors exhibit multiple-band superconducting gapping [50,52] and the enhancement of the transition temperature [52-59] when the condition of $2b < \xi_c(0)$ [52] is satisfied. The first discovered material in this class of superconductors is atomically thin FeSe [53-55] in which 13-fold increase (i.e., 100 K vs 7.5 K) was experimentally registered to date. Another milestone experimental finding in this field was reported by Liao *et al.* [9] who observed the effect of new superconducting band opening and $T_c$ enhancement in few layer stanene (which is the closest counterpart of graphene) by tuning the films thicknesses. To date, maximal $T_c$ increase due to the effect [52], stands with another single-atomic layer superconductor, $T_d$-MoTe$_2$, for which Rhodes *et al.* [59] reported 30-fold $T_c$ increase when samples were thinning down to single atomic layer.

In this paper we report results of our analysis of temperature dependent self-field critical currents, $I_c(sf,T)$, in Nb/BiSbTeSe$_2$-nanoribbon/Nb [48] and Nb/(Bi$_{0.06}$Sb$_{0.94}$)$_2$Te$_3$/Nb [60] junctions, and of the upper critical field, $B_{c2}(T)$, in Sn/SLG/Sn junctions [61] and show that a new superconducting band opening phenomenon in atomically thin superconductors, which we proposed earlier [50,52], has got further experimental supports.

## 2. Models description

In our previous work [50], we proposed to substitute $\Delta(T)$ in Eq. 1 by analytical expression proposed by Gross *et al.* [62]:

$$\Delta(T) = \Delta(0) \tanh\left(\frac{\pi k_B T_c}{\Delta(0)} \sqrt{\eta \left(\frac{\Delta C}{C}\right)\left(\frac{T_c}{T} - 1\right)}\right), \qquad (2)$$

where $\Delta(0)$ is the ground-state amplitude of the superconducting band, $\Delta C/C$ is the relative jump in electronic specific heat at the transition temperature, $T_c$, and $\eta = 2/3$ for *s*-wave superconductors [62].

In result, $T_c$, $\Delta C/C$, $\Delta(0)$, and normal-state tunneling resistance, $R_n$, of the S/N/S junction can be deduced by fitting experimental $I_c(sf,T)$ dataset to Eq. 1 (full expression for Eq. 1 is given in Ref. 50).

In Ref. 50 we showed that S/SLG/S and S/Bi$_2$Se$_3$/S junctions exhibit two-decoupled band superconducting state, for which, for general case of multiple-decoupled bands, temperature-dependent self-field critical current, $I_c(sf,T)$, can be described by the equation:

$$I_c(sf,T) = \sum_{i=1}^{N} \frac{\pi \cdot \Delta_i(T)}{2 \cdot e \cdot R_{n,i}} \cdot \theta(T_{c,i} - T) \cdot \tanh\left(\frac{\Delta_i(T)}{2 \cdot k_B \cdot T}\right), \qquad (3)$$

where the subscript $i$ refers to the $i$-band, $\theta(x)$ is the Heaviside step function, and each band has its own independent parameters of $T_{c,i}$, $\Delta C_i/C_i$, $\Delta_i(0)$, and $R_{n,i}$.

We should note that multiple-band induced superconductivity in junctions should be detectable by any technique which is sensitive to additional bands crossing the Fermi surface, for instance multiple distinct gaps should be evident in the temperature-dependence of the upper critical field, $B_{c2}(T)$, for which general equation is:

$$B_{c2}(T) = \sum_{i=1}^{N} B_{c2,i}(T) \cdot \theta(T_{c,i} - T), \qquad (4)$$

where, within each $i$-band, the upper critical field can be described by known model. In our study, we utilize four $B_{c2}(T)$ model to show that main result is model-independent. For instance, we use:

1. Two-fluid Gorter-Casimir (GC) model [63,64]:

$$B_{c2}(T) = \sum_{i=1}^{N}\left[B_{c2,i}(0) \cdot \left(1 - \left(\frac{T}{T_{c,i}}\right)^2\right) \cdot \theta(T_{c,i} - T)\right] = \frac{\phi_0}{2 \cdot \pi} \cdot \sum_{i=1}^{N}\left[\frac{\theta(T_{c,i} - T)}{\xi_i^2(0)} \cdot \left(1 - \left(\frac{T}{T_{c,i}}\right)^2\right)\right], \qquad (5)$$

where $\phi_0 = 2.068 \cdot 10^{-15}$ Wb is flux quantum, $\xi_i(0)$ is the ground state in-plane coherence length of the $i$-band. This model is a wide use for single-band superconductors ranging from 3D near-room-temperature superconducting hydrides [65-68] to 2D superconductors [54,55,61,69].

2. Jones-Hulm-Chandrasekhar (JHC) model [70]:

$$B_{c2}(T) = \frac{\phi_0}{2 \cdot \pi} \cdot \sum_{i=1}^{N} \frac{\theta(T_{c,i} - T)}{\xi_i^2(0)} \cdot \left(\frac{1 - \left(\frac{T}{T_{c,i}}\right)^2}{1 + \left(\frac{T}{T_{c,i}}\right)^2}\right), \qquad (6)$$

3. Werthamer-Helfand-Hohenberg model [71,72], for which we use analytical expression given by Baumgartner *et al.* [73] (we will designate this model as B-WHH herein):

$$B_{c2}(T) = \frac{\phi_0}{2 \cdot \pi} \cdot \sum_{i=1}^{N} \frac{\theta(T_{c,i} - T)}{\xi_i^2(0)} \cdot \left(\frac{\left(1 - \frac{T}{T_{c,i}}\right) - 0.153 \cdot \left(1 - \frac{T}{T_{c,i}}\right)^2 - 0.152 \cdot \left(1 - \frac{T}{T_{c,i}}\right)^4}{0.693}\right), \qquad (7)$$

4. Gor'kov model [74], for which simple analytical expression was given by Jones *et al.* [70]:

$$B_{c2}(T) = \frac{\phi_0}{2 \cdot \pi} \cdot \sum_{i=1}^{N} \frac{\theta(T_{c,i} - T)}{\xi_i^2(0)} \cdot \left(\left(\frac{1.77 - 0.43 \cdot \left(\frac{T}{T_{c,i}}\right)^2 + 0.07 \cdot \left(\frac{T}{T_{c,i}}\right)^4}{1.77}\right) \cdot \left[1 - \left(\frac{T}{T_{c,i}}\right)^2\right]\right), \qquad (8)$$

## 3. Results

### 3.1. Planar Sn/SLG/Sn array

Superconductivity in planar graphene junctions is varying by the change of the charge carrier density by moving away from the Dirac point in the dispersion [12,13,17]. This change is usually controlled by the gate voltage, $V_g$, applying to the junction. Han et al [61] reported on a proximity-coupled array of Sn discs with diameter of 400 nm on SLG which were placed in a hexagonal lattice separated by 1 μm between disks centers.

In Figs. 1 and 2 we show reported $B_{c2}(T)$ for Sn/SLG/Sn array by Han *et al.* [61] in their Figs. 4,5 at gate voltage of $V_g$ = 30 V. We defined $B_{c2}(T)$ by two criteria of $R$ = 0.01 kΩ (Fig. 1) and R = 0.2 kΩ (Fig. 2). It can be seen that there is an obvious upturn in $B_{c2}(T)$ at $T \leq 0.4 \cdot T_c$ independent of the upper critical field definition criterion. We note, that the upturn occurs at practically the same reduced temperature at which Borzenets *et al.* [15] observed the $I_c$(sf,$T$) enhancement in MoRe/SLG/MoRe junctions.

Accordingly, we fit these $B_{c2}(T)$ datasets to four two-band models (Eqs. 5-8) and ones are shown in Figs. 1,2. Deduced parameters, including the ratio of transition temperatures for two bands, $\frac{T_{c2}}{T_{c1}} = 0.32 \pm 0.02$ for $R$ = 0.01 kΩ criterion (Fig. 1), and $\frac{T_{c2}}{T_{c1}} = 0.38 \pm 0.01$ for $R$ = 0.2 kΩ criterion (Fig. 2), are well agreed with each other despite a fact that experimental $B_{c2}(T)$ data were processed by four different models.

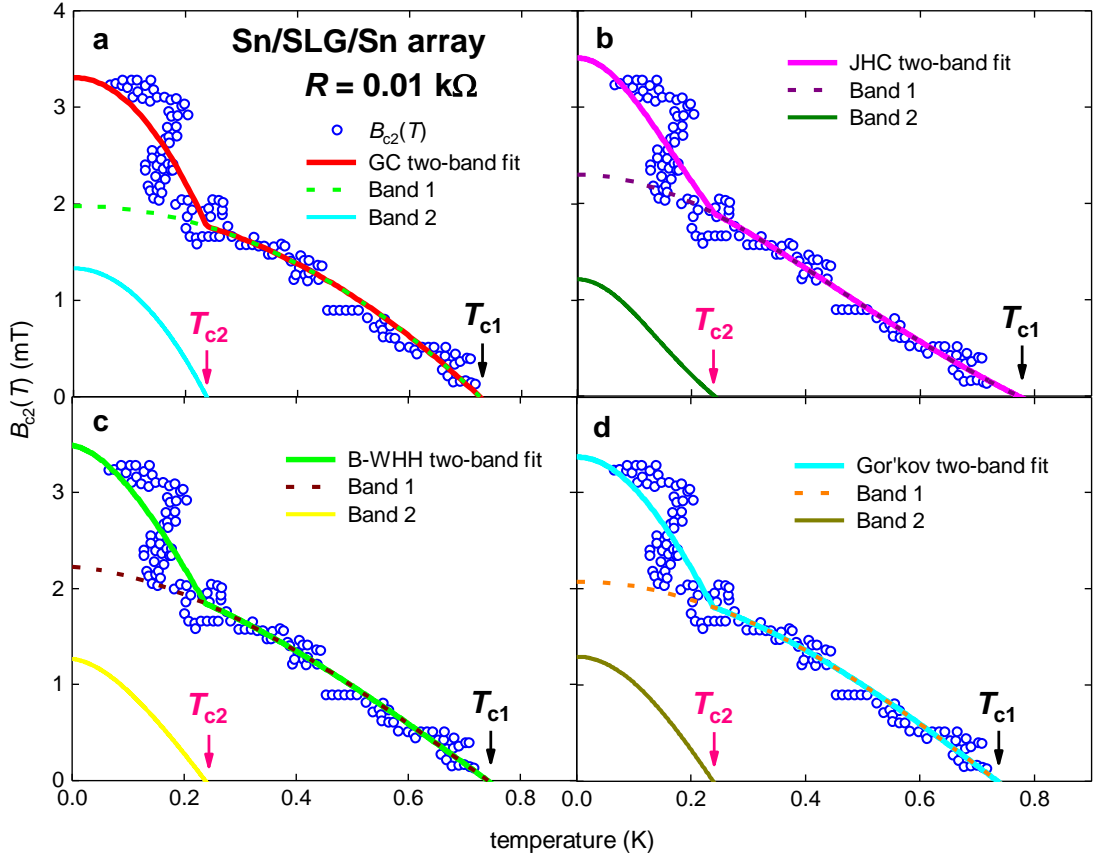

**Figure 1.** Experimental $B_{c2}(T)$ for Sn/SLG/Sn array at gate voltage of $V_g$ = 30 V [61] and data fits to Eqs. 5-8. $B_{c2}$ criterion is $R$ = 0.01 kΩ. (**a**) GC model. Derived parameters: $T_{c1}$ = 0.72 ± 0.01 K, $\xi_1(0)$ = 408 ± 7 nm, $T_{c2}$ = 0.24 ± 0.01 K, $\xi_2(0)$ = 497 ± 22 nm, $\frac{T_{c2}}{T_{c1}} = 0.33 \pm 0.02$, fit quality is $R$ = 0.9059; (**b**) JHC model. Derived parameters: $T_{c1}$ = 0.77 ± 0.02 K, $\xi_1(0)$ = 378 ± 8 nm, $T_{c2}$ = 0.24 ± 0.02 K, $\xi_2(0)$ = 521 ± 34 nm, $\frac{T_{c2}}{T_{c1}} = 0.31 \pm 0.04$, fit quality is $R$ = 0.9101; (**c**) B-WHH model. Derived parameters: $T_{c1}$ = 0.74 ± 0.02 K, $\xi_1(0)$ = 385 ± 7 nm, $T_{c2}$ = 0.24 ± 0.01 K, $\xi_2(0)$ = 510 ± 28 nm, $\frac{T_{c2}}{T_{c1}} = 0.32 \pm 0.02$, fit quality is $R$ = 0.9093. (**d**) Gor'kov model. Derived parameters: $T_{c1}$ = 0.74 ± 0.02 K, $\xi_1(0)$ = 398 ± 7 nm, $T_{c2}$ = 0.24 ± 0.01 K, $\xi_2(0)$ = 504 ± 25 nm, $\frac{T_{c2}}{T_{c1}} = 0.32 \pm 0.02$, fit quality is $R$ = 0.9082.

We also need to note that experimental data of Han *et al.* [61] have an evidence that there is the third upturn in $B_{c2}(T)$ which can be seen at lowest experimentally available temperatures of $T$ < 0.1 K and applied fields of about $B$ ~ 4.5 mT in Fig. 4,5 [61], if the criterion of $R$ ~ 0.05 kΩ (for the $B_{c2}(T)$ definition) will be applied.

Despite a fact that authors [61] did not mention the presence of these two upturns in raw experimental $B_{c2}(T)$ data and more detailed measurements of $B_{c2}(T)$ requires to reveal more accurately the position and parameters for the third band, there is already enough experimental evidences that Sn/SLG/Sn array exhibits at least two-superconducting bands gapping, and thus, the report of Han *et al.* [61] supports our primary idea that atomically thin films exhibits multiple-band superconducting gapping phenomenon [50,52].

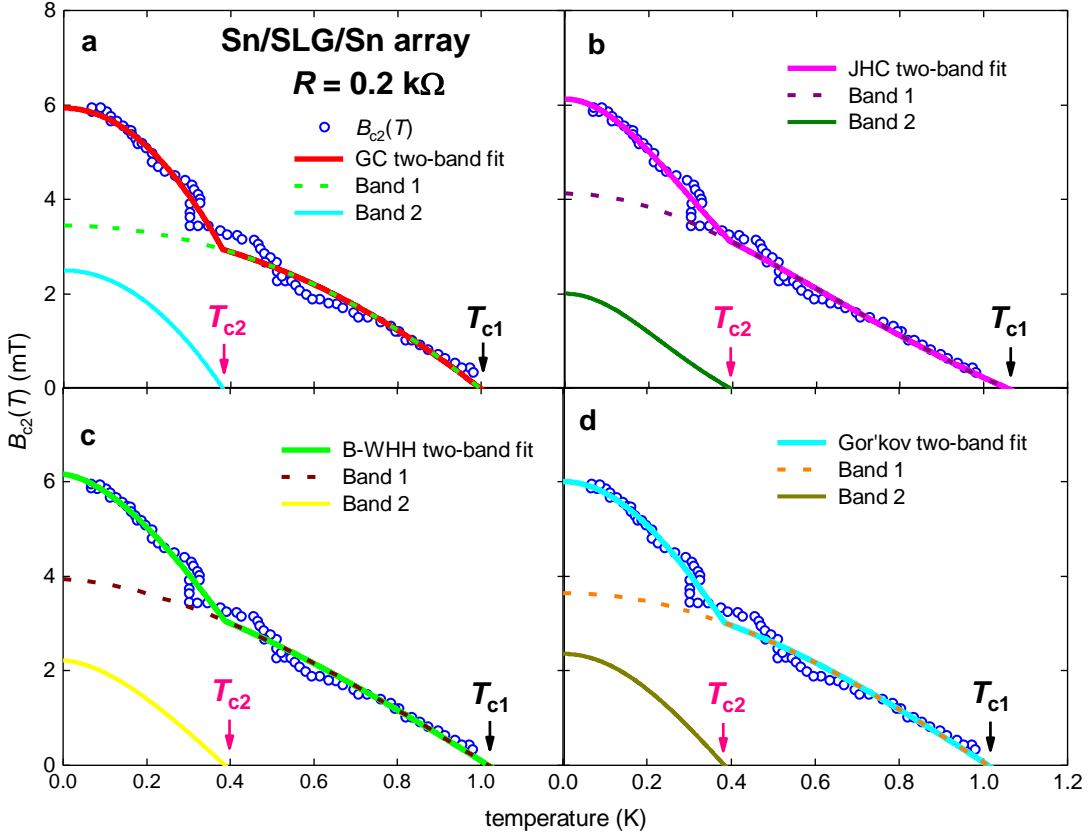

**Figure 2.** Experimental $B_{c2}(T)$ for Sn/SLG/Sn array at gate voltage of $V_g$ = 30 V [61] and data fits to Eqs. 5-8. $B_{c2}$ criterion is $R$ = 0.2 kΩ. (**a**) GC model. Derived parameters: $T_{c1}$ = 1.00 ± 0.01 K, $\xi_1(0)$ = 309 ± 3 nm, $T_{c2}$ = 0.38 ± 0.01 K, $\xi_2(0)$ = 363 ± 7 nm, $\frac{T_{c2}}{T_{c1}}$ = 0.38 ± 0.01, fit quality is $R$ = 0.9847; (**b**) JHC model. Derived parameters: $T_{c1}$ = 1.06 ± 0.01 K, $\xi_1(0)$ = 283 ± 3 nm, $T_{c2}$ = 0.39 ± 0.01 K, $\xi_2(0)$ = 405 ± 11 nm, $\frac{T_{c2}}{T_{c1}}$ = 0.37 ± 0.01, fit quality is $R$ = 0.9903; (**c**) B-WHH model. Derived parameters: $T_{c1}$ = 1.02 ± 0.01 K, $\xi_1(0)$ = 289 ± 3 nm, $T_{c2}$ = 0.39 ± 0.01 K, $\xi_2(0)$ = 385 ± 9 nm, $\frac{T_{c2}}{T_{c1}}$ = 0.38 ± 0.01, fit quality is $R$ = 0.9885. (**d**) Gor'kov model. Derived parameters: $T_{c1}$ = 1.01 ± 0.01 K, $\xi_1(0)$ = 300 ± 3 nm, $T_{c2}$ = 0.38 ± 0.01 K, $\xi_2(0)$ = 374 ± 7 nm, $\frac{T_{c2}}{T_{c1}}$ = 0.38 ± 0.01, fit quality is $R$ = 0.9873.

## 3.2. Planar Nb/BiSbTeSe$_2$-nanoribbon/Nb junctions

There is a wide accepted view that the superconducting state in S/TI/S junctions, similarly to the case of S/SLG/S junctions, controls by the gate voltage, $V_g$. In this regard, junctions made of the tetradymite compound, BiSbTeSe$_2$, one of the most bulk-insulating three-dimensional topological insulators [47], should follow these expectations. However, our analysis of recent experimental data reported by Kayyalha *et al.* [48] on Nb/BiSbTeSe$_2$/Nb junctions, shows that superconducting state in BiSbTeSe$_2$-based systems is very robust vs the change in the gate voltage, $V_g$, and thus, at least, S/TI/S junctions where TI thickness is less than the ground state of the coherence length, $\xi(0)$, have different physical operation principles than S/SLG/S counterparts. For instance, Kayyalha *et al.* [48] in their Figs. 2 and S1 reported $I_c(\text{sf},T)$ for five Nb/BiSbTeSe$_2$-nanopribbon/Nb junctions at different $V_g$. The thickness of BiSbTeSe$_2$ flakes was varied from $2b$ = 30 nm to 50 nm, and based on reported $\xi(0)$ ~ 600 nm [48], the condition of $2b < \xi(0)$ [50,52] is satisfied for all junctions.

### 3.2.1. Nb/BiSbTeSe$_2$-nanoribbon/Nb junctions (Sample 1 [48])

In Fig. 3 we show experimental $I_c(sf,T)$ datasets for Sample 1 [48] reported for three gate voltages, $V_g$ = -20 V (Fig. 3,a), 0 V (Fig. 3,b), and +45 V (Fig. 3,c). $I_c(sf,T)$ fits to Eq. 3 were performed for all parameters to be free as experimental raw datasets were rich enough to carry out this sort of fits.

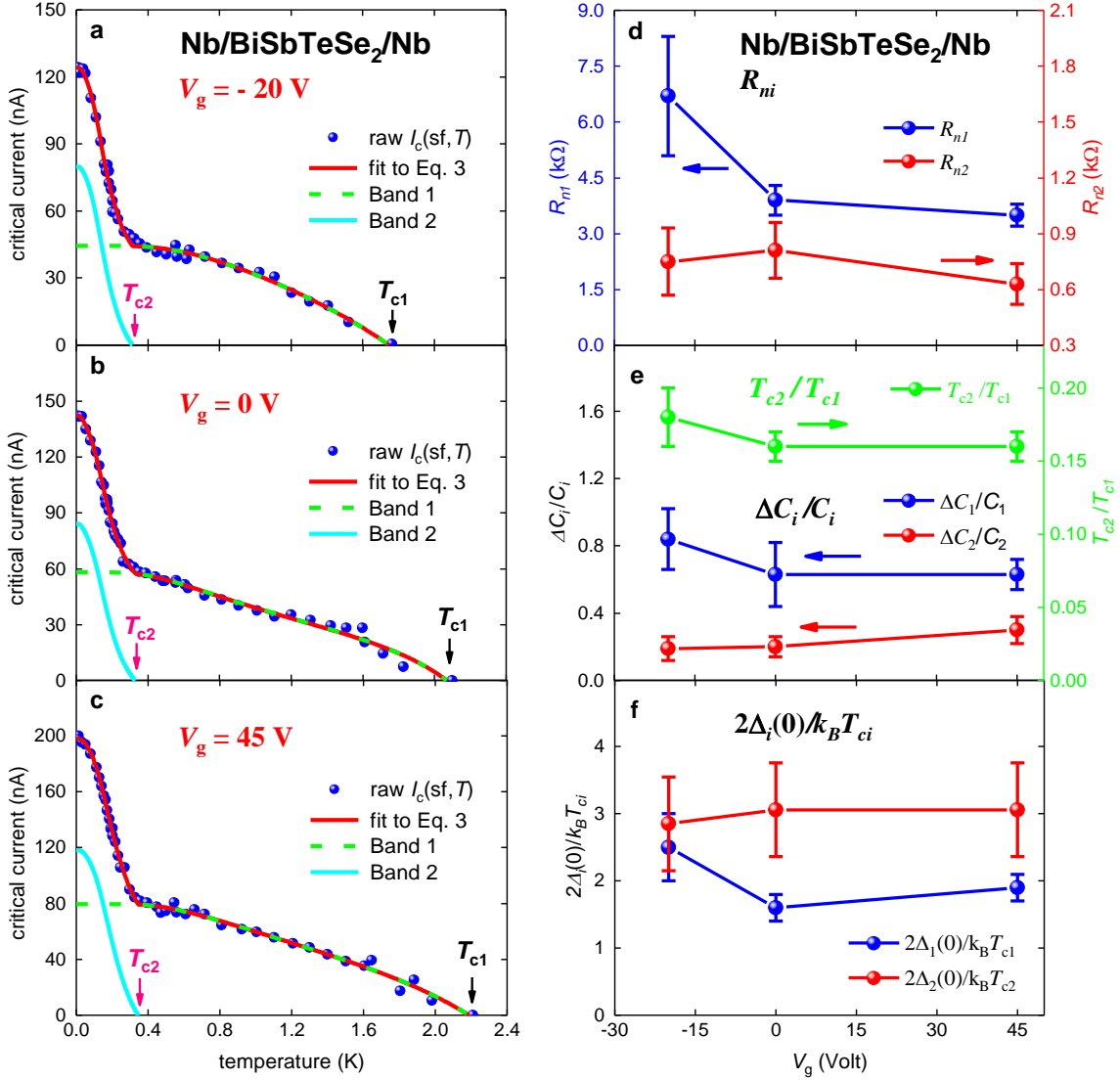

**Figure 3.** Experimental $I_c(sf,T)$ for Nb/BiSbTeSe$_2$-nanoribbon/Nb junction (Sample 1 [48]), data fits to Eq. 3, and major deduced parameters. (**a**) Gate voltage $V_g$ = - 20 V. Derived parameters: $T_{c1}$ = 1.74 ± 0.04 K, $\Delta_1(0)$ = 190 ± 40 µeV, $\Delta C_1/C_1$ = 0.84 ± 0.18, $2\Delta_1(0)/k_BT_{c1}$ = 2.5 ± 0.5, $R_{n1}$ = 6.7 ± 1.6 kΩ, $T_{c2}$ = 0.31 ± 0.02 K, $\Delta_2(0)$ = 38.2 ± 9.7 µeV, $\Delta C_2/C_2$ = 0.19 ± 0.07, $2\Delta_2(0)/k_BT_{c2}$ = 2.85 ± 0.70, $R_{n2}$ = 0.75 ± 0.18 kΩ, $\frac{T_{c2}}{T_{c1}}$ = 0.18 ± 0.02, fit quality is $R$ = 0.9953; (**b**) Gate voltage $V_g$ = 0 V. Derived parameters: $T_{c1}$ = 2.07 ± 0.03 K, $\Delta_1(0)$ = 144 ± 11 µeV, $\Delta C_1/C_1$ = 0.63 ± 0.19, $2\Delta_1(0)/k_BT_{c1}$ = 1.6 ± 0.2, $R_{n1}$ = 3.9 ± 0.4 kΩ, $T_{c2}$ = 0.33 ± 0.02 K, $\Delta_2(0)$ = 43.5 ± 8.4 µeV, $\Delta C_2/C_2$ = 0.20 ± 0.06, $2\Delta_2(0)/k_BT_{c2}$ = 3.06 ± 0.70, $R_{n2}$ = 0.81 ± 0.15 kΩ, $\frac{T_{c2}}{T_{c1}}$ = 0.16 ± 0.01, fit quality is $R$ = 0.9965; (**c**) Gate voltage $V_g$ = 45 V. Derived parameters: $T_{c1}$ = 2.19 ± 0.03 K, $\Delta_1(0)$ = 176 ± 13 µeV, $\Delta C_1/C_1$ = 0.63 ± 0.09, $2\Delta_1(0)/k_BT_{c1}$ = 1.9 ± 0.2, $R_{n1}$ = 3.5 ± 0.3 kΩ, $T_{c2}$ = 0.34 ± 0.01 K, $\Delta_2(0)$ = 47.6 ± 8.7 µeV, $\Delta C_2/C_2$ = 0.30 ± 0.08, $2\Delta_2(0)/k_BT_{c2}$ = 3.06 ± 0.70, $R_{n2}$ = 0.63 ± 0.11 kΩ, $\frac{T_{c2}}{T_{c1}}$ = 0.16 ± 0.01, fit quality is $R$ = 0.9977; (**d**) Derived $R_{ni}$ as function of gate voltage $V_g$; (**e**) Derived $\frac{T_{c2}}{T_{c1}}$ and $\Delta C_i/C_i$ as function of gate voltage $V_g$; (**f**) Derived $2\Delta_i(0)/k_BT_{ci}$ as function of gate voltage $V_g$.

As the result, we deduced $R_{ni}$, $\frac{T_{c1}}{T_{c2}}$, $\Delta C_i/C_i$, $\Delta_i(0)$, and $\frac{2 \cdot \Delta_i(0)}{k_B \cdot T_{c,i}}$ for both superconducting bands as functions of applied gate voltage, $V_g$. These deduced parameters are shown in Figs. 3 (e-f).

We need to stress, that within the range of uncertainties, deduced $R_{n1}$ values are well agree with directly measured values by Kayyalha *et al.* [48] (these values reported in Fig. 1 (a) of Ref. 48). More often measured raw $I_c(sf,T)$ data, and especially at high reduced temperatures, are required to reduce the uncertainty for $R_{n1}$ values.

Most notable outcome of our analysis is that, within uncertainty ranges, fundamental superconducting parameters for both bands, including the ratio of $\frac{T_{c2}}{T_{c1}}$, are remaining unchanged vs gate voltage variation in the range from -10 V to 45 V. This means that two-band superconducting state in Nb/BiSbTeSe$_2$-nanoribbon/Nb junction is very robust and mostly independent from the change in the gate voltage, $V_g$. This is unexpected result, because there is generally accepted view that because gate voltage, $V_g$, is determined the electronic state in 2D-systems in the normal state, ones should also determine the superconducting state. However, our analysis shows that this is not a case in general view. As we already mentioned above, there is a need for more often measurements of raw $I_c(sf,T)$ data, which will allows to reduce uncertainties for all deduced parameters.

3.2.2. Nb/BiSbTeSe$_2$-nanoribbon/Nb (Sample 3 [48])

In Fig. 4 (a) we show experimental $I_c(sf,T)$ dataset for Nb/BiSbTeSe$_2$-nanoribbon/Nb (Sample 3) reported by Kayyalha *et al.* [48].

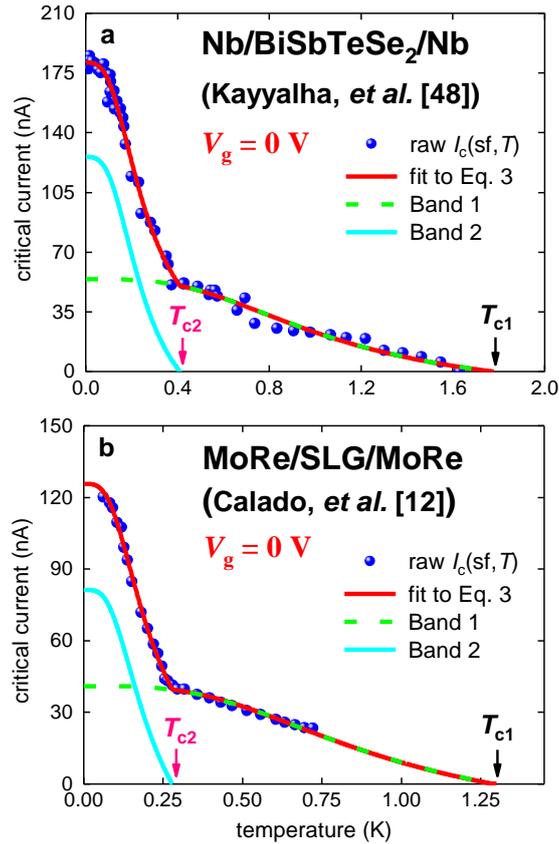

**Figure 4.** Experimental $I_c(sf,T)$ for two atomically thin DCM-based junctions and fits to Eqs. 3,9,10. (**a**) Nb/BiSbTeSe$_2$/Nb (Sample 3 [48]). Derived parameters: $T_{c1}$ = 1.8 ± 0.1 K, $\Delta_1(0)$ = 179 ± 51 μeV, $\Delta C/C$ = 0.20 ± 0.04, $2\Delta(0)/k_BT_c$ = 2.3 ± 0.7, $R_{n1}$ = 5.2 ± 1.4 kΩ, $T_{c2}$ = 0.41 ± 0.02 K, $\Delta_2(0)$ = 41 ± 12 μeV, $R_{n2}$ = 0.51 ± 0.15 kΩ, $\frac{T_{c2}}{T_{c1}}$ = 0.23 ± 0.02, $R$ = 0.9954; (**b**) MoRe/SLG/MoRe (Sample A [12]). Derived parameters: $T_{c1}$ = 1.29 ± 0.07 K, $\Delta_1(0)$ = 139 ± 36 μeV, $\Delta C/C$ = 0.30 ± 0.04, $2\Delta(0)/k_BT_c$ = 2.5 ± 0.7, $R_{n1}$ = 5.3 ± 1.4 kΩ, $T_{c2}$ = 0.28 ± 0.01 K, $\Delta_2(0)$ = 30 ± 8 μeV, $R_{n2}$ = 0.56 ± 0.16 kΩ, $\frac{T_{c2}}{T_{c1}}$ = 0.22 ± 0.01, $R$ = 0.9981.

Raw experimental $I_c(sf,T)$ dataset for this sample was not reach enough at $T \geq 0.6\,K$, and thus we cannot perform the fit to Eq. 3 for all parameters to be free. To run the model (Eq. 3), we make the same model restriction, as we did in our previous work [50]:

$$\frac{\Delta C_1}{C_1} = \frac{\Delta C_2}{C_2} = \frac{\Delta C}{C}, \quad (9)$$

$$\frac{2\Delta_1(0)}{k_B \cdot T_1} = \frac{2\Delta_2(0)}{k_B \cdot T_2} = \frac{2\Delta(0)}{k_B \cdot T_c}, \quad (10)$$

i.e., we forced $\Delta C_i/C_i$ and $\frac{2 \cdot \Delta_i(0)}{k_B \cdot T_{c,i}}$ values to be the same for both bands. As the result, we deduce $R_{ni}$, $T_{ci}$, $\frac{T_{c2}}{T_{c1}} \sim \frac{1}{4}$, $\Delta C/C$, $\Delta_i(0)$, and $\frac{2\cdot\Delta(0)}{k_B \cdot T_c}$ for this junction and find that these values are very close to ones deduced for Sample 1 (Fig. 3).

*3.3. Planar MoRe/SLG/MoRe junction (Device A [12])*

To demonstrate that our findings in regard of Nb/BiSbTeSe$_2$-nanoribbon/Nb junctions are generic for a much wide range of atomically-thin DCM-based Josephson junctions, in Fig. 4 (b) we show raw $I_c(sf,T)$ dataset and fit to our model (Eq. 3) for MoRe/SLG/MoRe reported by Calado *et al.* [12] for their Device A [12]. For the $I_c(sf,T)$ fit for this device we used the same parameters restrictions (Eqs. 9,10), as for Nb/BiSbTeSe$_2$-nanoribbon/Nb Sample 3 [48].

In our previous work [50], we already analyzed this $I_c(sf,T)$ dataset for MoRe/SLG/MoRe Device A [12]. However, what was a surprise, that there is remarkable and practically undistinguishable similarity in reduced $I_c(sf,T)$ datasets and fits for Nb/BiSbTeSe$_2$-nanoribbon/Nb [48] and MoRe/SLG/MoRe [12] junctions (Fig. 4). In attempt to make further extension for our findings belong S/DCM/S junctions, in next Section we analyze $I_c(sf,T)$ data for Nb/(Bi$_{0.06}$Sb$_{0.94}$)$_2$Te$_3$-nanoribbon/Nb junction [58].

*3.4. Planar Nb/(Bi$_{0.06}$Sb$_{0.94}$)$_2$Te$_3$-nanoribbon/Nb junction*

In Fig. 5 we show temperature-dependent self-field critical currents, $I_c(sf,T)$, in Nb/(Bi$_{0.06}$Sb$_{0.94}$)$_2$Te$_3$-nanoribbon/Nb reported by Schüffelgen *et al.* [58], where TI nanoribbon has thickness of $2b$ = 10 nm, and, thus, the condition of $2b < \xi(0)$ [50,52] was satisfied.

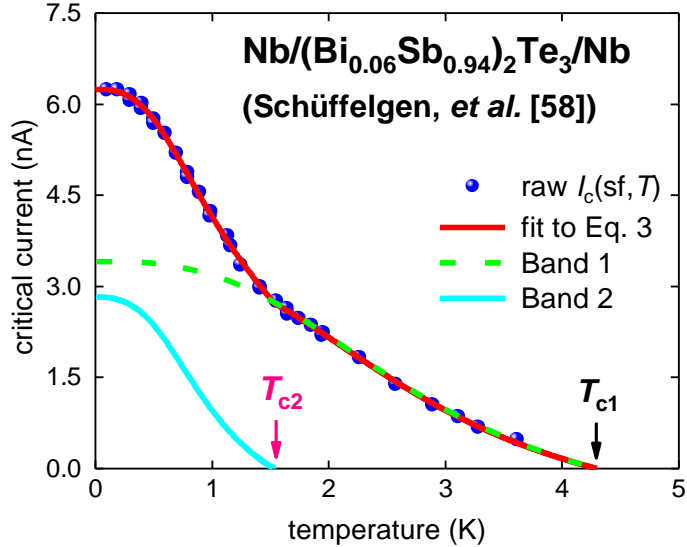

**Figure 5.** Experimental $I_c(sf,T)$ for atomically thin DCM-based junction Nb/(Bi$_{0.06}$Sb$_{0.94}$)$_2$Te$_3$-nanoribbon/Nb [58] and fit to Eq. 3,9,10. Derived parameters: $T_{c1}$ = 4.30 ± 0.07 K, $\Delta_1(0)$ = 530 ± 7 μeV, $\Delta C/C$ = 0.28 ± 0.04, $2\Delta(0)/k_BT_c$ = 2.87 ± 0.05, $R_{n1}$ = 244 ± 32 Ω, $T_{c2}$ = 1.53 ± 0.03 K, $\Delta_2(0)$ = 189 ± 3 μeV, $R_{n2}$ = 105 ± 16 Ω, $\frac{T_{c2}}{T_{c1}}$ = 0.36 ± 0.01, $R$ = 0.9995.

Due to reported $I_c$(sf,$T$) dataset was not rich enough at high reduced temperatures, we restrict model by utilizing Eqs. 9,10. In overall, fitted curves and all deduced parameters are very close to one reported by Borzenets *et al.* [15] for MoRe/SLG/MoRe junctions (which we processed and showed in our previous paper [50] in Fig. 7).

## 4. Discussion

We should stress that Calado *et al.* [12] in 2015 requested the necessity for a new model to explain the upturn in $I_c$(sf,$T$) registered in their MoRe/SLG/MoRe junction (Device A) at $T \sim \frac{1}{4} \cdot T_c$ (which we show in Fig. 4 (b)), because this $I_c$(sf,$T$) enhancement was not possible to explain neither by Eilenberger model (which is in use to describe clean S/N/S junctions) [75], nor by Isadel model (which describes diffusive S/N/S junctions) [76].

Our explanation for this upturn [52], which is well aligned with the $I_c$(sf,$T$) upturn in natural atomically thin superconductors [50], is that this $I_c$(sf,$T$) enhancement is due to a new superconducting band opening phenomenon when sample dimensions become smaller than some critical value. For this critical value we proposed to use [52] the out-of-plane coherence length, $\xi_c(0)$, which is still, after expanding our analysis herein, a good choice for the scaling criterion.

We need pointed out that this new opening band phenomenon is not necessarily causes the increase in observed transition temperature in comparison with "bulk" material. For instance, in pure Nb films [77], this new "thin film" band has lower transition temperature in comparison with "bulk" band [52]. And when this is the case, there is no warning for the researcher to search more deeply created device/films for new superconducting band.

Thus, perhaps, in many atomically thin films, which in fact exhibit a new band opening phenomenon, this effect was not registered yet, because there was no expectation that something important/interesting can be observed at low reduced temperatures, well below "bulk" or observed $T_c$ for given atomically thin film.

We also need to note, that the effect of new superconducting band opening [52] in atomically thin films can be detected by any experimental techniques which is sensitive to additional band(s) crossing the Fermi surface. To date, most evident confirmations for the phenomenon are related to the $I_c$(sf,$T$) upturn [9,50,52] and $B_{c2}(T)$ upturn [9], however, other techniques also should detect this.

In this regard, we want to mention non-ambitiously observation of the $I_c$(sf,$T$) upturn reported by Li *et al.* [44] in their Fig. 4 (a) at $T$ = 2.5 K in Nb/Cd$_3$As$_2$-nanowire/Nb junction. However, raw experimental $I_c$(sf,$T$) dataset [44] was limited by measurements at $T$ < 3.5 K, and thus, we are not able to perform the analysis for this very interesting atomically-narrow S/TI/S junction at the moment.

There are very interesting results reported by Sasaki *et al.* [78] and by Andersen *et al.* [79], who found that temperature-dependent upper critical field, $B_{c2}(T)$, in nanostructures of topological insulators, cannot be explained by single-band WHH model [71,72]. However, reported, to date, raw experimental $B_{c2}(T)$ datasets [77,78] are not reach enough to perform two-band model fit to reveal the presence of additional band at low reduced temperatures in these structures.

We also need to mention an interesting research field of interfaced superconductivity [37,80,81], where, as we proposed earlier [52], the effect of new band opening [52] should play major role. However, the discussion of this interesting field is beyond the scope of this paper.

## 5. Conclusions

As the result, in this paper we perform analysis of recently reported experimental data on induced superconducting state in atomically thin Dirac-cone films. We show, that the phenomenon of new superconducting band opening in atomically thin films [50,52], when the film thickness becomes thinner than the ground state out-of-plane coherence length, $\xi_c(0)$, can be extended on induced superconducting state in atomically thin DCM, as one was established before for natural superconductors, i.e. pure Nb, exfoliated 2H-TaS$_2$, double-atomic layer FeSe and few layer stanene [9].

**Funding:** This research was funded by the State Assignment of Minobrnauki of Russia, theme "Pressure" No. AAAA-A18-118020190104-3, and by Act 211 Government of the Russian Federation, contract No. 02.A03.21.0006.

**Acknowledgments:** Author would like to thank Dr. S. Goswami and Prof. L. M. K. Vandersypen (Kavli Institute of Nanoscience, Delft University of Technology, The Netherlands) for providing raw self-field critical current data for the MoRe/SLG/MoRe devices analyzed in this work.

**Conflicts of Interest:** The funders had no role in the design of the study, in the collection, analyses, or interpretation of data, in the writing of the manuscript, or in the decision to publish the results.